   \documentclass[aps,prb,preprint,groupedaddress,endfloats]{revtex4}
   \usepackage{graphicx}
   \usepackage{amssymb}
   \usepackage{amsmath}
   \usepackage{subfigure}
   \usepackage{graphpap}
   \usepackage{dcolumn} 
   \usepackage{bm}
   \usepackage{color,calc}
   \usepackage{longtable}
\begin{document}
   \preprint{}
   \title{Second-Order Nonlinear Optics from First Principles}
   \author{Eleonora \surname{Luppi}, Hannes \surname{H\"ubener}, Val\'erie \surname{V\'eniard}}
   \affiliation{Laboratoire des Solides Irradi\'es,\'Ecole Polytechnique, CNRS-CEA/DSM, 
                F-91128 Palaiseau, \\
                European Theoretical Spectroscopy Facility (ETSF), France}
   \date{\today}
   \begin{abstract}
   We present a new first-principle theory for the calculation of the 
   macroscopic second-order susceptibility $\chi^{(2)}$, based 
   on the Time-Dependent Density-Functional Theory approach.
   Our method allows to include straightforwardly the many-body effects.
   We apply the theory to the computation of the 
   Second-Harmonic Generation spectroscopy, showing
   a very good agreement with experiment for cubic semiconductor GaAs.
    \end{abstract}
   \pacs{42.65.An 42.70.Mp 71.15.Mb 78.20.-e}
   \maketitle
\begin{figure}
   \includegraphics[width=6.0cm,angle=-90,clip]{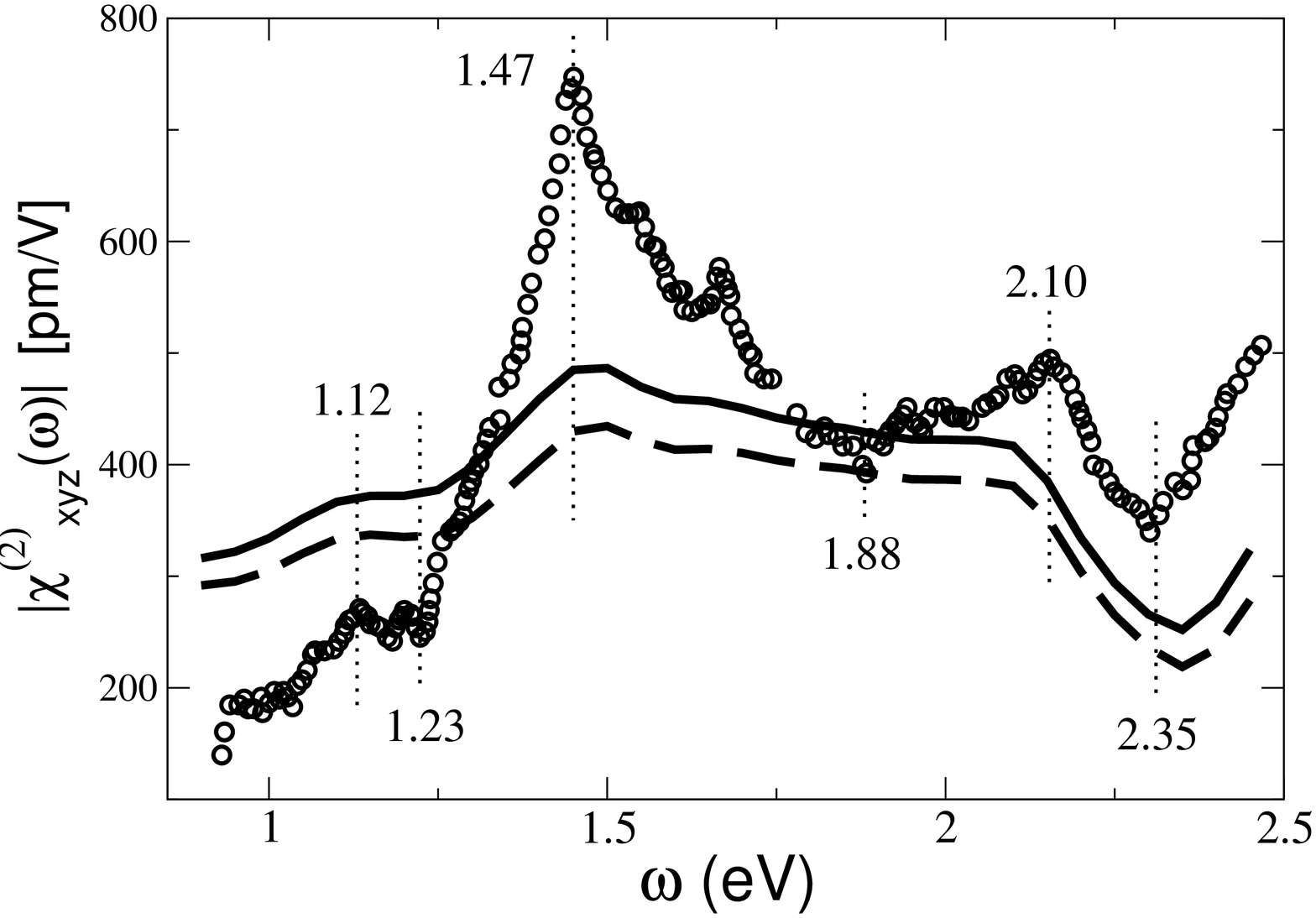}
   \caption{\label{GaAs_iparpa} $\vert\chi^{(2)}_{xyz}\vert$ calculated within 
           IPA (solid line) and including crystal local fields (dashed line).
           Comparison with the most accurate experimental data available now \cite{berg-daum03prl} (circle).
           The experimental energy position of peaks and valleys are reported.}
   \end{figure}

   Nonlinear optics is one of the most important and exciting field of fundamental and applied research,
   with applications in physics, chemistry and biology.
   Among the nonlinear phenomena existing in nature, the main role is played by second-order processes, like 
   Second-Harmonic Generation (SHG) and Sum-Frequency Generation (SFG), which 
   are extremely versatile tools for studies of many kinds of surfaces 
   and interfaces \cite{shen98nat}. 
   Their interest is rapidly growing, because of their exceptional sensitivity to space symmetry violations, and
   nowadays SHG and SFG are also used
   for characterizing systems like nanocrystal interfaces \cite{figl+05prl} or as a probe
   for molecular chirality in polymer \cite{oh-e+04prl} and
   nanotubes \cite{su+08prb}.
   The inverse process of SFG: Parametric Amplification together with
   Optical Rectification are used in microwave and terahertz technology \cite{tono07nat}.\\
   In all these processes, the interaction of matter with light is described by
   the macroscopic second-order susceptibility $\chi^{(2)}$.
   In principle, $\chi^{(2)}$ includes the many-body interactions among the electrons of the system:
   the variation of the screening fields on the microscopic scale, i.e. crystal local-field effects \cite{hybe-loui87prb}
   and  the electron-hole interaction, i.e. excitonic effects \cite{onid+02rmp}, as real and/or virtual excitations are created
   in the process.
   The theoretical description of the many-body effects in the second-order response is 
   a big challenge and only a small number of \emph{ab initio} works exist on the topic, mainly focused 
   on the static or low energy limit.
   Self-consistent local-field effects were included in $\chi^{(2)}$ within local density approximation (LDA)
   through the ``2n+1'' theorem applied to the action functional as defined in 
   Time-Dependent Density-Functional Theory (TDDFT) \cite{dalc+96prb} or in a band theory \cite{levi94prb,chen+97prb} 
   in the case of semiconductor 
   and insulator materials. 
   Electron-hole interaction has been described, in the second-order response, through the solution of an effective
   two-particle Hamiltonian, derived in the Bethe-Salpeter equation approach (BSE) \cite{chan+01prb,leit+05prb}.
   However, the validity of this approach has been demonstrated in linear-response calculations and 
   the question arises whether it is possible to use this method also
   for higher-order calculations.
   This indicates the crucial need for different many-body approaches.\\ 
   In this letter we present a new first-principles theory 
   for the calculation of the static and dynamic macroscopic second-order
   susceptibility $\chi^{(2)}$, based on the Time-Dependent Density-Functional
   Theory (TDDFT) approach \cite{rung-gros84prl,gros-kohn85prl}.
   This formulation is valid for any kind of crystals: semiconductors and metals.
   The goal of our formalism is twofold:
   first, the exact relation (non-relativistic regime) between {\it microscopic} and {\it macroscopic}
   formulation of the second-order response, 
   shown here for the first time, and second, a rigorous and straightforward 
   treatment of the many-body effects.   
   In order to validate our theory we have applied our method to the SHG spectroscopy
   for the frequently studied material: cubic semiconductor GaAs.
   Indeed, GaAs has been object of large interest in SHG since its discovery, and a 
   certain number of experimental and theoretical studies
   have been performed,  but no existing  theoretical approaches 
   have been able to give a  conclusive comparison with
   experiments.   
   We are able to do it with our formalism,  showing here for the first time, 
   an  excellent agreement with experimental data.\\
   The objective of our theory is to find an expression for the susceptibility  
   $\chi^{(2)}$, defined through the macroscopic second-order polarization
   \begin{equation}
   P^{(2)}_{M} = \chi^{(2)}E^{tot}E^{tot}, 
   \label{polmacro2}
   \end{equation}
   $E^{tot}$ being the macroscopic component of the total electric-field.
   $E^{tot}$ includes the contribution from the external perturbing electric-field
   and from the electric-field due to the polarization of the system, 
   induced by the external perturbation. \\
   Our starting point is the calculation of the second-order \textit{microscopic} polarization 
   via the second-order time-dependent perturbation theory.
   We obtain an expression for the \textit{microscopic} polarization in term of the
   perturbing electric-field $E^{p}$
   \begin{eqnarray}
   P^{(2)}({\bf q}+{\bf G},\omega)=\sum_{{ \bf q}_1,{ \bf q}_2,{\bf G}_1, {\bf G}_2 }
                            \int d \omega_1  d \omega_2 
   \delta({\bf q} - { \bf q}_1 - { \bf q}_2)  
   \nonumber \\[0.3cm]
   \times \delta(\omega -  \omega_1 - \omega_2 ) \tilde{\alpha}^{(2)}
   ({\bf q}+{\bf G},{ \bf q}_1 + {\bf G}_1,{ \bf q}_2 + {\bf G}_2,\omega_1,\omega_2)
   \nonumber \\[0.4cm]
   \times E^{p}({ \bf q}_1 + {\bf G}_1,\omega_1) E^{p}({ \bf q}_2 + {\bf G}_2,\omega_2), \hspace{2.91cm}
   \label{pmicro}
   \end{eqnarray}
   where $\tilde{\alpha}^{(2)}$ is the quadratic quasi-polarizability tensor.
   All quantities are functions of the frequency $\omega$, of the vectors ${\bf q}$ 
   in the Brillouin zone and of the reciprocal lattice vectors ${\bf G}$.\\
   In the first step of our calculation, all 
   the quantities are written in term of the perturbing electric-field and are  \textit{microscopic}.
   At this point, there are two main difficulties in our formalism: first we 
   need to average correctly in space to obtain macroscopic
   measurable quantities and second we have to express the polarization as a function of the
   total-electric field.\\
   To overcome these problems, we introduce the function
   \begin{equation}
   F({\bf q},\omega) = \left[ 1+ 4\pi \frac{\tilde{\alpha}^{(1)}(\bf{q},\bf{q},\bf{\omega})}
                                   {1-4\pi \alpha^{(1),LL}(\bf{q},\bf{q},\bf{\omega})} \right]
   \end{equation}
   were $\tilde{\alpha}^{(1)}$ is the linear quasi-polarizability and $\alpha^{(1),LL}$ is its longitudinal-longitudinal
   contraction, and we use, in Eq.~(\ref{pmicro}) the relation between the perturbing and the total electric field (macroscopic component)
   \begin{eqnarray}
   E^{p}({\bf q},\omega)=F({\bf q},\omega) E^{tot}({\bf q},\omega),
   \label{epetot}
   \end{eqnarray}
   demonstrated by Del Sole and Fiorino \cite{dels-fior84prb}.\\
   We thus obtain
   the desired results: the \textit{macroscopic} second-order polarization
   \begin{eqnarray}
   P^{(2)}_{M}({\bf q},\omega) = \sum_{{ \bf q}_1,{ \bf q}_2,}  \int d \omega_1  d \omega_2
                                 \delta({\bf q} - { \bf q}_1 - { \bf q}_2)
                                 \nonumber \\[0.3cm]
                                 \times \delta(\omega -  \omega_1 - \omega_2 )F({\bf q}) 
                                 \tilde{\alpha}^{2}({\bf q},{ \bf q}_1,{ \bf q}_2,\omega_1,\omega_2) 
                                 \nonumber \\[0.4cm]
                                 \times F( {\bf q}_1) F({ \bf q}_2)
                                 E^{tot}({\bf q}_1,\omega_2) E^{tot}({ \bf q}_2,\omega_2), \hspace{0.5cm}
   \label{pmacro}
   \end{eqnarray}
   from which it is easy to derive, through a comparison 
   with  Eq.~(\ref{polmacro2}), our key quantity, the susceptibility $\chi^{(2)}$.
   $P^{(2)}_{M}$ (Eq.~(\ref{pmacro})) contains the \emph{ab initio} relation between the \textit{microscopic} and \textit{macroscopic} formulation of the second-order response,
   and it is shown here for the first time.\\
   Furthermore, our formalism is completely general for electric fields containing both longitudinal and 
   transverse components. 
   In the following, we will consider only the case 
   of vanishing light wave vector ($q \rightarrow 0$), for which longitudinal and transverse responses are equal, since the direction of $\mathbf q$ is no longer defined 
   for a uniform field and the responses depend only on the polarization of the field \cite{pinenozibook}.
   Therefore, we have expressed the second-order susceptibility in terms of longitudinal quantities only.\\
   We have derived the optical susceptibility for any crystal symmetry and here we show the case of the cubic zinc-blend symmetry which 
   has only one independent non-vanishing component (out of 18 tensor components)
   \begin{eqnarray}
   \chi_{xyz}^{(2)}(\omega_{1}+\omega_{2},\omega_{1},\omega_{2})=  -\frac{i}{12}\lim _{{\bf q}\to 0}
   \frac{1}{{\bf q}_{x}{\bf q}_{y}{\bf q}_{z}}
   \nonumber \\[0.35cm]   
   \times \chi_{\rho\rho\rho}({\bf q},{\bf q},\omega_{1},\omega_{2})
   \epsilon^{LL}_M(\omega_{1}+\omega_{2})
   \epsilon^{LL}_M(\omega_{1}) \epsilon^{LL}_M(\omega_{2}),
   \label{chixyz}
   \end{eqnarray}
   where ${\bf q}=({\bf q}_{x},{\bf q}_{y},{\bf q}_{z})$.
   In this scalar equation,
   $\epsilon^{LL}_M$ is the macroscopic longitudinal dielectric function,  
   defined as $D=\epsilon^{LL}_M E_{tot}$  (for longitudinal fields) and
   it has to be evaluated at the photon frequencies $\omega_{1}$, $\omega_{2}$ and $\omega_{1}+\omega_{2}$.\\
   The second-order susceptibility $\chi^{(2)}_{xyz}$ depends also on the $\chi_{\rho\rho\rho}$, which is 
   the longitudinal second-order response function. 
   All these quantities are evaluated in the limit ${\bf q}\to 0$.
   The general expression for the $\chi_{\rho\rho\rho}$, within TDDFT, is given through the generalized
   matrix Dyson equation
   \begin{eqnarray}
   \left[ 1-\chi_0^{(1)}(\omega_1 + \omega_2) f_{uxc}(\omega_1 + \omega_2)\right]\chi^{(2)}_{\rho\rho\rho}(\omega_1,\omega_2) =  
   \nonumber\\[0.3cm]
   \chi_0^{(2)}(\omega_1,\omega_2) 
   \left[1 + f_{uxc}(\omega_1) \chi^{(1)}(\omega_1)\right] \hspace{1.2cm}
   \nonumber\\[0.3cm]
   \times \left[1 + f_{uxc}(\omega_2)\chi^{(1)}(\omega_2)\right] \hspace{1.2cm}
   \nonumber\\[0.3cm]
   +\,\,\chi_0^{(1)}(\omega_1 + \omega_2)g_{xc}(\omega_1 + \omega_2)\chi^{(1)}(\omega_1)\chi^{(1)}(\omega_2) 
   \label{dyson2}
   \end{eqnarray}
   where $f_{uxc}$ is the sum of the bare-coulomb potential $u$ and of the exchange-correlation kernel $f_{xc}=\frac{\delta V_{xc}}{\delta \rho}$.
   A new kernel, $g_{xc}=\frac{\delta^2 V_{xc}}{\delta \rho \delta \rho'}$, appears,
   defined as the second derivative of the exchange-correlation potential. 
   Moreover, $\chi^{(1)}(\omega)$ is the linear response function calculated via the Dyson equation 
   \begin{equation}
   [1 - \chi^{(1)}_{0}(\omega)f_{uxc}(\omega)]\chi^{(1)}(\omega) = \chi^{(1)}_{0}(\omega). \label{dyson1}
   \end{equation}
   The functions $\chi^{(1)}_{0}(\omega)$ and $\chi^{(2)}_{0}(\omega)$ are the linear and second-order response functions in the
   Independent-Particle Approximation (IPA).
   The response function $\chi_{\rho\rho\rho}$ can be calculated with different levels of approximation, depending on the kernels we use. 
   Up to now, most of the \emph{ab initio} calculations existing in literature were obtained within IPA, which we recover by 
   setting $f_{xc}=0$ and $g_{xc}=0$. 
   In this case the factors $1+u\chi^{(1)}$ and $1-\chi_{0}^{(1)}u$ of Eq.~(\ref{dyson2}) compensate the $\epsilon^{LL}_M$ functions of Eq.~(\ref{chixyz}),
   leading to the usual expression $\chi_{xyz}^{(2)}=\chi^{(2)}_{0}$.
   Note that in Eq.~(\ref{dyson2}) and Eq.~(\ref{dyson1}) we have omitted the explicit dependence on the $\bf{q}$ and $\bf{G}$-vectors 
   of the response functions.\\ 
   To corroborate our theory, we show SHG spectra for cubic semiconductor GaAs,
   which is one of the most studied system in nonlinear optics from theory and experiments. 
   We compare with the most accurate experimental data available now, obtained by Bergfeld and Daum \cite{berg-daum03prl}, 
   who determined
   the $\vert\chi^{(2)}_{xyz}\vert$ ( reported in Fig.~(\ref{GaAs_iparpa})) 
   over a wide spectral range.
  The absolute value of the susceptibility $\vert\chi^{(2)}_{xyz}\vert$ has been determined 
   by measurements of the reflected second-harmonic light
   (note that the presence of the surface leads to additional peaks that cannot be separated from the bulk contribution E=1.65 eV).
   In their work \cite{berg-daum03prl}, two theoretical studies are mentioned \cite{hugh-sipe96prb,leit+05prb}, showing only a qualitative agreement with experiment,
   and not in the whole frequency range.
   Furthermore, none of them is able to reproduce the absolute values,
   even if many-body effects, in the framework of the BSE, are considered \cite{leit+05prb}.\\
   We have performed our electronic ground-state calculations with the plane-waves pseudopotential code ABINIT \cite{Abinit}
   and employed, for the nonlinear optics, a new nonlinear-response code implemented by us, 
   on the basis of the linear-response Dp code \cite{Dp}.
   One of the crucial steps for the estimation of the second-order response function
   is the computation of the band structure of the system. 
   For Gallium, Dal Corso {\it et al.}  \cite{dalc+96prb} demonstrate the importance of the 3d semicore states,
   pointing out that to obtain
   correct $\vert\chi^{(2)}_{xyz}\vert$  values it is necessary to include
   at least the nonlinear core correction to the Gallium pseudopotential.
   In our calculations we have explicitly included the  3$d$ semicore states of Gallium
   as valence states.
   In fact, the presence or the absence of the 3$d$ semicore states
   influences the value of $\vert\chi^{(2)}_{xyz}\vert$ for GaAs, 
   resulting in a slightly increasing of the  $\vert\chi^{(2)}_{xyz}\vert$ when including the $d$ states.\\
   \begin{figure}
   \includegraphics[width=6.0cm,angle=-90,clip]{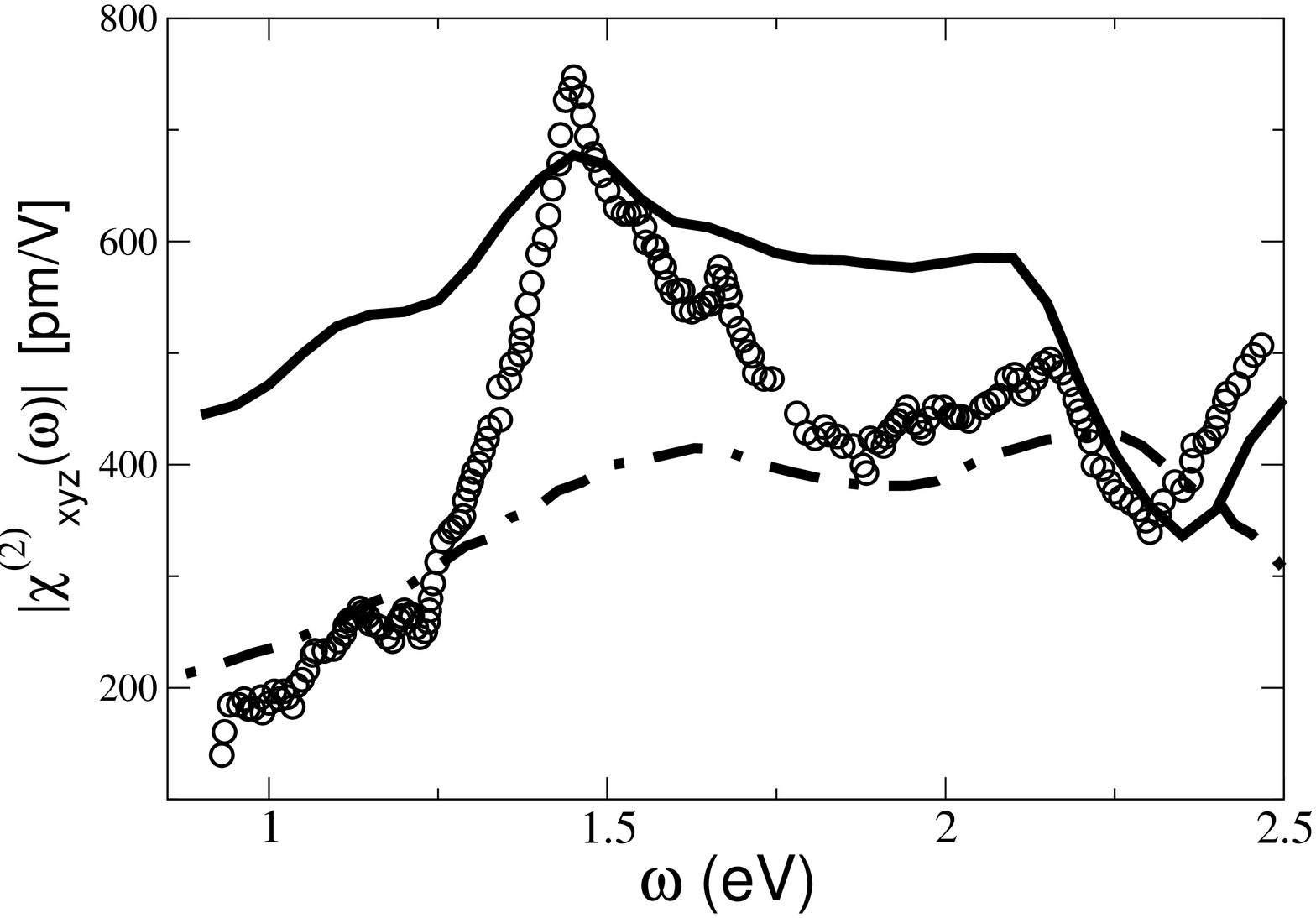}
   \caption{\label{GaAs_alphaexp}
           The experimental $\vert\chi^{(2)}_{xyz}\vert$ for GaAs (circle) \cite{berg-daum03prl} is compared to our calculation (solid line), which includes the
           excitonic effects within TDDFT through the $\alpha$ kernel and with the calculation of 
           Leitsmann {\emph et al.} \cite{leit+05prb} (dot-dashed line)
           where the excitons are included within BSE framework.}
   \end{figure}
   In  Fig.~(\ref{GaAs_iparpa}), together with the experimental result \cite{berg-daum03prl}, is also shown the 
   calculated $\vert\chi^{(2)}_{xyz}\vert$ in the independent-particle approximation and 
   with the inclusion of  the crystal local-field effects.   
   In both cases we have
   correctly \cite{nast+05prb} applied the scissor shift \cite{dels-girl93prb} of 0.8 eV to the Kohn-Sham band structure.
   Already within IPA we obtain the same shape as the experimental spectra. 
   Both the theoretical and the experimental spectrum have three peaks and three valleys which are almost
   at the same energy position.
   This agreement can be achieved only when including explicitly the semicore states in the calculation.
   Taking into account the crystal local fields, for this energy range, the main effect is a
   decrease of the magnitude of the second-order susceptibility of the order of 10\%, when compared to IPA.
   The shape of the spectrum is not significantly changed.
   However, even though the shape of the theoretical spectrum is good, 
   the relative intensity of the peaks and in particular the magnitude of the susceptibility is not in agreement with the 
   experimental values.
   The physics of the process is not sufficiently described neither in the independent-particle picture nor
   taking into account the microscopic inhomogeneity of the system.\\
   To go beyond these approximations, we have considered the excitonic effects through the $f_{xc}$ kernel, keeping $g_{xc}=0$. 
   In Eq.~(\ref{chixyz}), $f_{xc}$
   appears in the calculation of $\epsilon_{M}$, $\chi^{(1)}$
   and $\chi_{\rho\rho\rho}$.
   When using the ALDA kernel (not shown here) for $f_{xc}$
   the result remains very close to those of IPA and IPA with crystal local fields.
   This is very similar to the failure of TDLDA absorption spectra in solids \cite{gavr-bech97prb}, related to the lack of  long-range contribution in the 
   ALDA kernel.
   To solve this issue, a model long-range kernel of the form $\alpha/q^2$
   has been proposed \cite{rein+02prl},
   where
   $\alpha$ is a mean value for the dynamical dependence of $f_{xc}$,
   in a given range of frequency.
   For GaAs the standard values for $\alpha$ are 0.05 in the static limit and
   0.2 in the energy range of Fig.~(\ref{GaAs_alphaexp}).
   The main effect
   of the $\alpha/q^2$ kernel is to increase the magnitude of $\vert\chi^{(2)}_{xyz}\vert$ 
   recovering the order of magnitude of the absolute value of the experimental second-order susceptibility without
   changing the position of the energy peaks and valleys, as shown in Fig.~(\ref{GaAs_alphaexp}).
   This behavior can be understood by solving analytically Eq.~(\ref{dyson2}) without local fields,
   showing that the increase from the $\vert\chi^{(2)}_{0}\vert$ to the $\vert\chi^{(2)}\vert$ is
   proportional to
   $[1+\alpha/4\pi(\epsilon^{LL}_M(\omega)-1)]^2[1+\alpha/4\pi(\epsilon^{LL}_M(2\omega) -1)]$.\\
   Only excitonic effects can correctly describe the magnitude of SHG measurements in GaAs.
   They have to be included carefully and consistently with a second-order process.
   In fact, in Fig.~(\ref{GaAs_alphaexp}) it is also reported the spectrum calculated by Leitsmann {\emph et al.} \cite{leit+05prb} where
   excitons are described in the second-order susceptibility using the BSE approach. 
   The $\vert\chi^{(2)}_{xyz}\vert$ is much lower than the experiments.
   We believe that in the nonlinear response regime, for finite frequencies,
   the crystal local fields and the excitons,
   are not correctly described by this effective Hamiltonian derived in the BSE approach.\\
   Instead, in the static limit, these effects seem to be less important and this BSE-based method is still valid.
   Indeed, we obtain 205 pm/V for $\vert\chi_{xyz}^{(2)}\vert$ which is in agreement with Chang {\emph et al.} who obtained 236.4 pm/V and
   both results are in 
   a reasonable agreement with experimental data: 180$\pm$10 pm/V \cite{levi-beth72apl} and 166 pm/V \cite{robe92ieee} at frequency 0.117 eV 
   and 172 pm/V at frequency 0.118 eV \cite{eyre+01prb}.\\
   In conclusion, in this letter, we have presented a new first-principles formalism for the
   calculation of the static and dynamic second-order susceptibility  $\chi^{(2)}$, in the 
   Time-Dependent Density-Functional Theory framework, valid for any type of crystal (semiconductor and metal).
   Our theory permits to write down, for the first time, the exact relation (non-relativistic regime) 
   between {\it microscopic} and {\it macroscopic}
   formulation of the second-order response function, together with a 
   rigorous and straightforward 
   treatment of the many-body effects.   
   We have applied this formalism to the most accurate experimental result made on cubic semiconductor GaAs,
   giving, for the first time, conclusive theoretical explanation of the experimental data.
   We show that only carefully including excitonic effects, it is possible to recover the magnitude
   of the experimental $\chi^{(2)}$.
   Finally, we want to emphasize the importance of our formalism.
   This theory represents crucial progress for the development of nonlinear optics, which is 
   a  very active and exciting field in many disciplines
   of research and we
   are convinced that this work can open new ways to explore it.\\
   We wish to thank L. Reining, Ch. Giorgetti, F. Bechstedt, A. Rubio  and F. Sottile for helpful discussions. 
   This work was supported by the European Union through the NANOQUANTA Network of Excellence and the 
   ETSF Integrated Infrastructure Initiave, as well as by the ANR project ANR-05-BLANC-0191-01-LSI 171.
   Computations have been performed at the Institut du D\'{e}veloppement et des Ressources en Informatique 
   Scientifique (IDRIS) and at the 
   Centre de Calcul Recherche et Technologie (CCRT).

   \end{document}